# Rapid oxygen exchange between hematite and water vapor

**Authors**

Zdenek Jakub[1]†, Matthias Meier[1,2], Florian Kraushofer[1], Jan Balajka[1], Jiri Pavelec[1], Michael Schmid[1], Cesare Franchini[2,3], Ulrike Diebold[1], Gareth S. Parkinson[1*]

**Affiliations**

[1]Insitute of Applied Physics, TU Wien, Vienna, Austria

[2]University of Vienna, Faculty of Physics and Center for Computational Materials Science, Vienna, Austria

[3] Alma Mater Studiorum - Università di Bologna, Bologna, Italy

†current affiliation: Central European Institute of Technology (CEITEC), Brno University of Technology, Czech Republic

*correspondence to: parkinson@iap.tuwien.ac.at

**Abstract**

Oxygen exchange at oxide/liquid and oxide/gas interfaces is important in technology and environmental studies, as it is closely linked to both catalytic activity and material degradation. The atomic-scale details are mostly unknown, however, and are often ascribed to poorly defined defects in the crystal lattice. Here we show that even thermodynamically stable, well-ordered surfaces can be surprisingly reactive. Specifically, we show that all the 3-fold coordinated lattice oxygen atoms on a defect-free single-crystalline "r-cut" ($1\bar{1}02$) surface of hematite ($\alpha$-$Fe_2O_3$) are exchanged with oxygen from surrounding water vapor within minutes at temperatures below 70 °C, while the atomic-scale surface structure is unperturbed by the process. A similar behavior is observed after liquid water exposure, but the experimental data clearly show most of the exchange happens during desorption of the final monolayer, not during immersion. Density functional theory computations show that the exchange can happen during on-surface diffusion, where the cost of the lattice oxygen extraction is compensated by the stability of an HO-HOH-OH complex. Such insights into lattice oxygen stability are highly relevant for many research fields ranging from catalysis and hydrogen production to geochemistry and paleoclimatology.

**Introduction**

Atom exchange phenomena at solid/gas or solid/liquid interfaces are crucial for a wide range of fields including catalysis (1-4), energy storage (2, 4-6), geochemistry (7-9), corrosion studies (10, 11) or paleoclimatology (7, 12). In heterogeneous catalysis, atom exchange is an integral part of Mars-van-Krevelen-type mechanisms, where the product contains an atom removed from the surface (3, 13). The catalytic cycle is closed by replenishing the surface atom vacancy from the gas or liquid phase, and if this happens significantly slower than the product formation, the catalyst eventually degrades. Without the knowledge of the atomic-scale mechanism, an experimental observation of atom exchange can thus indicate higher activity, faster degradation, or both (2, 4, 14).

The most detailed views on atom exchange phenomena come from experiments carried out in well-defined, highly idealized conditions which allow for precise computational modelling: on monodisperse nanometer-sized clusters in liquid (11, 15, 16), or at extended surfaces in ultrahigh vacuum (UHV) (13, 17). The work on nanometer-sized clusters primarily focuses on environmentally relevant water/solid oxygen exchange, and reveals that exchange mechanisms can be highly complex and counter-intuitive. Many atoms are involved in a single exchange event, and the exchange rate at 2-fold coordinated ($\mu_2$-O) sites can sometimes be faster than at 1-fold coordinated ($\eta$=O) sites. Nevertheless, in all studied cases the exchange rates at a 3- ($\mu_3$-O) or 4-fold ($\mu_4$-O) coordinated sites were several orders of magnitude slower, often comparable to the dissolution rate of the whole cluster.

Surface science observations of atom exchange processes are common in catalysis-oriented studies (13, 17, 18), but relatively rare for environmentally relevant processes (19, 20). Water/surface oxygen exchange is often observed at point defects such as surface oxygen vacancies (19, 21, 22), but such defects are seldom found in ambient. On presumably defect-free oxide surfaces, observations of extensive oxygen exchange are scarce (19, 20, 23), and attempts to elucidate the mechanism are hindered by poor knowledge of the surface structures (11, 19, 23). To date, the most convincing case has been reported on the $(1\bar{1}02)$ surface of hematite ($\alpha$-$Fe_2O_3$) (20), where the atomic-scale structure of the surface has since been confirmed by multiple experimental techniques both in vacuum and in liquid water (24-26). The nature of the facile oxygen exchange remains puzzling, however, as there is currently no acceptable explanation for a mechanism involving at minimum three Fe-O bond ruptures (exchange at a $\mu_3$-O site) to be preferred over a simple desorption of a water molecule into vacuum (11). If it turns out such phenomena are common on surfaces, identification of the structural parameters facilitating rapid atom exchange can lead to efficient strategies to enhance it (e.g., in catalysis and energy storage), to prevent it (in corrosion protection), or to estimate its extent (in geochemistry or paleoclimatology).

In this work we elucidate the water/surface oxygen exchange on $\alpha$-$Fe_2O_3(1\bar{1}02)$, which is a dominant facet on both natural hematite and synthetic nanomaterial. We address the following questions: In which conditions does the rapid oxygen exchange take place, how the mechanism proceeds at the atomic scale, and what are the potential implications for applied fields. Using Temperature Programmed Desorption (TPD), Low-Energy He$^+$ Ion Scattering (LEIS) and X-Ray Photoemission Spectroscopy (XPS), we find that the oxygen exchange mechanism happens rapidly and continuously when the surface is exposed to low pressure of water vapor at temperatures between 30-70 °C. In the presence of $10^{-8}$ mbar $H_2O$ vapor it only takes minutes to completely exchange the whole top layer of stable 3-fold coordinated O atoms. This contrasts with the observations after immersion to liquid water and exposures to near-ambient pressure water vapor at room temperature, where the amount of exchanged oxygen is lower, comparable to that observed after a single TPD experiment in UHV. The experimental data are rationalized by density functional theory (DFT) computations, which show that the lattice oxygen exchange can happen during surface diffusion of $H_2O$. This is because the energetic cost of a lattice oxygen atom extraction is partially compensated by formation of a cooperatively-stabilized HO-HOH-OH complex. The diffusion process inherently requires unoccupied cation sites on the surface, and thus it happens sluggishly in ambient or liquid, but rapidly in vacuum. These results yield atomic-level insights into the complex dynamics of stable mineral surfaces and provide a clear example of local water-water interactions defining the chemistry of oxide surfaces.

**Results**

**Experiments in ultrahigh vacuum**

The ($1\bar{1}02$) surface of a natural hematite (α-$Fe_2O_3$) single crystal represents an ideal model system to study mineral surface chemistry because a well-defined bulk-truncated surface can be easily prepared in UHV by sputtering (1 keV $Ar^+$ or $Ne^+$, 10 min) and annealing in a partial pressure of $O_2$ (10-20 min, 476 °C, $P_{O2}$ = 5×$10^{-7}$ mbar) (20, 24, 25, 27). The surface has been studied in detail previously (20, 24, 27, 28), and its atomic structure can be described as zig-zag rows of 3-fold coordinated surface oxygen atoms running in the ($1\bar{1}0\bar{1}$) direction. These are straddled by 5-fold coordinated Fe(III) cations (Fig. 1a). The stoichiometric, bulk-truncated surface is also stable in aqueous environment for at least several days (26), and it was recently shown that short water immersion does not induce any changes detectable by ambient Atomic Force Microscopy (AFM), XPS or Low Energy Electron Diffraction (LEED) (25). For completeness it is to be noted that this bulk-truncated surface termination differs from the one reported previously on samples prepared by chemical-mechanical polishing, which feature areas depleted of the topmost Fe layer (29-32). Upon annealing in air this Fe-deficient surface reverts to the bulk-truncated surface studied here (26, 33).

The atomic-scale structure of a water monolayer on α-$Fe_2O_3$($1\bar{1}02$) was previously shown to consist of partially-dissociated HO-$H_2O$ dimers (Fig. 1b,c) (25). In UHV, two distinct water phases can be formed on the surface depending on the water coverage and temperature. This results in two desorption peaks in $D_2O$ TPD spectra (Fig. 1d). When the water monolayer is complete (2 molecules/unit cell (u.c.), peak β in Fig. 1d), these HO-$H_2O$ dimers are densely packed and an $H_2O$ or OH completes the octahedral coordination of every surface Fe cation (Fig. 1b). At room temperature, the full water monolayer in UHV is unstable against desorption of 1/3 of the water, resulting in a lower-coverage phase (1.33 molecules/u.c., peak γ in Fig. 1d) which features an empty adsorption site next to each HO-$H_2O$ dimer in the ($1\bar{1}0\bar{1}$) direction (Fig. 1c). In addition to TPD experiments and DFT computations, these two water phases were unambiguously identified by atomic-scale non-contact Atomic Force Microscopy imaging (ncAFM) and XPS (25).

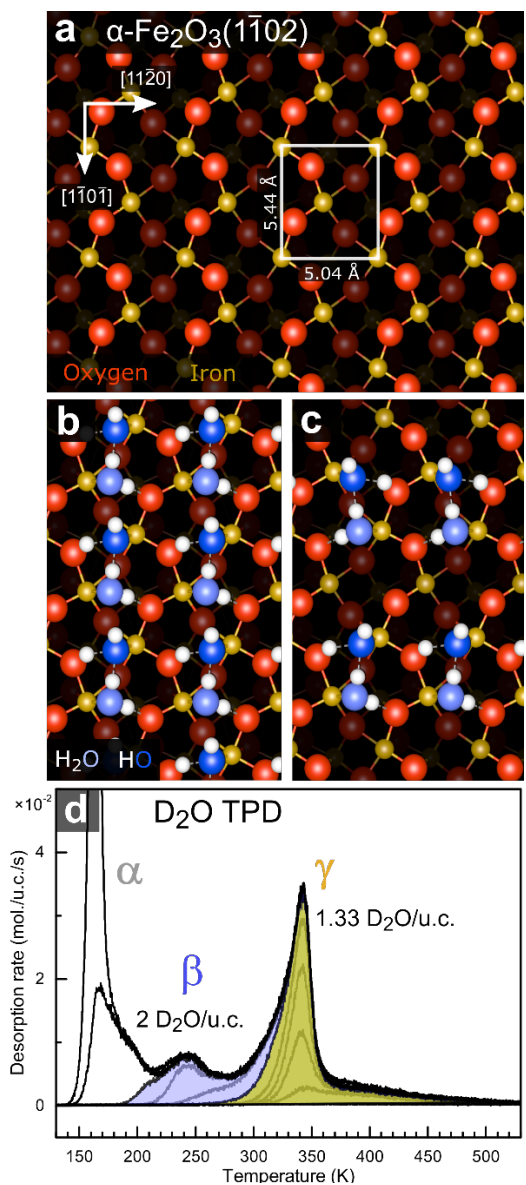

**Figure 1:** Water adsorption on α-$Fe_2O_3$(1$\bar{1}$02). (**a**) Top view on the bulk-truncated α-$Fe_2O_3$(1$\bar{1}$02) surface. The red (larger) and yellow (smaller) balls correspond to oxygen and iron; the surface unit cell contains two O and two Fe atoms. A full analysis of this surface termination is provided in reference (24). In this work, the azimuthal angle of 0° is defined as the (1$\bar{1}$0$\bar{1}$) direction. (**b,c**) Two phases of water adsorbed on this surface, both of which consist of partially-dissociated water dimers (25). The bright blue and dark blue balls correspond to oxygen in $H_2O$ and OH, respectively. (**d**) $D_2O$ TPD (temperature ramp 1 K/s) shows two main desorption peaks (β, γ) before the multilayer forms (α). The saturation coverages of the β, γ peaks correspond to models shown in B and C, respectively.

Figure 2a shows TPD measurements performed using water with isotopically labelled oxygen, $H_2^{18}O$. Here, the desorption signal from the full monolayer (peak β in Fig. 1d) is observed almost exclusively in the $m/e$ = 20 channel corresponding to $H_2^{18}O$, whereas desorption from the lower-coverage phase takes place through two channels of similar intensity, corresponding to $H_2^{18}O$ and $H_2^{16}O$ ($m/e$ = 20 and $m/e$ = 18; the $m/e$ = 18 spectrum shown is corrected for $H_2^{16}O$ adsorption from the background, the cracking pattern of the $H_2^{18}O$, and $H_2^{16}O$ impurity in the dosed $H_2^{18}O$, see Supplementary Note 1). This indicates facile oxygen exchange occurred between the dosed $H_2^{18}O$ and the α-$Fe_2^{16}O_3$ lattice. Moreover, the data imply a fundamental difference in how water desorbs from the full water monolayer and the more stable γ phase shown in Fig. 1c. The ratio of the $m/e$ = 20 and $m/e$ = 18 signal maxima amounts to 54/46, i.e., almost half of the water molecules in the γ desorption peak contains an oxygen atom extracted from the hematite lattice. Taking into account the ratio between the number of surface oxygen atoms (2 in a unit cell) and the number of oxygen atoms in the γ phase (1.33 in a unit cell), ≈ 30 % of the surface oxygen atoms on a perfect surface of an α-$Fe_2O_3$(1$\bar{1}$02) single crystal exchange in a single water adsorption-desorption cycle.

The oxygen exchange is also observed during continuous $H_2^{18}O$ exposure at 350 K (i.e. the temperature of the desorption rate maximum in TPD), as shown in Fig. 2b. Upon opening the shutter of the incident $H_2^{18}O$ molecular beam, a significant $H_2^{16}O$ signal is observed, with an initial intensity of ≈ 50% of the $H_2^{18}O$ signal. Within minutes, however, the $H_2^{16}O$ signal decreases as the $H_2^{18}O$ signal increases simultaneously; the overall water signal remains constant. This behavior is due to gradual $^{18}O$ enrichment of the α-$Fe_2^{16}O_3$(1$\bar{1}$02) surface. After circa 6 minutes (a dose corresponding to ≈ 10 $H_2^{18}O$/u.c.), almost all the water scattered from the surface is observed in the $m/e$ = 20 channel, indicating that all the surface O atoms available for exchange have been exchanged with the incoming water at least once.

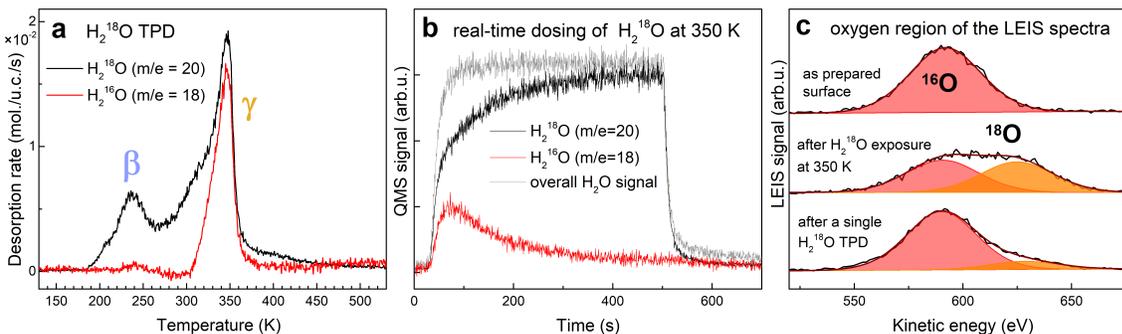

**Figure 2:** Observation of the oxygen exchange between $H_2^{18}O$ and α-$Fe_2^{16}O_3$ by TPD and LEIS. (**a**) TPD spectrum (1 K/s) after a dose of ≈ 1.9 $H_2^{18}O$/u.c. at 120 K. The β peak is observed mostly in the $m/e$ = 20 signal, while the γ peak desorbs in two channels of similar intensity, indicating substantial oxygen exchange with the α-$Fe_2^{16}O_3$ substrate. (**b**) Oxygen exchange observed during 2.7×10$^{−8}$ mbar $H_2^{18}O$ exposure at 350 K. Upon opening of the molecular beam shutter, roughly 1/3 of the signal is observed in the $m/e$ = 18 channel. After ca. 360 s, almost all the signal is contained in the $m/e$ = 20 channel due to the saturation of the surface with $^{18}O$. The overall water signal remains constant. (**c**) LEIS characterization (1 keV He$^+$, scattering angle 90°) of the α-$Fe_2^{16}O_3$ as prepared (top), after the $H_2^{18}O$ exposure as shown in B (middle) and after a single $H_2^{18}O$ TPD experiment (bottom). The latter two spectra show substantial enrichment of the α-$Fe_2^{16}O_3$ surface with $^{18}O$. Source data are provided as a Source Data file.

The TPD data are further supported by LEIS, which can efficiently distinguish between surface atoms with different masses, such as $^{16}O$ and $^{18}O$. On the as-prepared surface, the LEIS spectrum shows a single O peak that can be fitted well with a single symmetric pseudo-Voigt component (top spectrum in Fig. 2c). On a surface saturated with $^{18}O$ (as shown in Fig. 2b), the LEIS spectrum (middle in Fig. 2c) can be fitted by two O peaks with the same full width at half-maximum as the peak measured on the clean surface. The areas of the two fitted peaks are almost equivalent (ratio 51/49). This result is consistent with the whole top O layer being exchanged, because in this experimental setup the incoming (1 kV He$^+$) and outgoing ion direction is oriented along the rows of the surface (azimuthal angle 0°, polar angle 45°). Thus, it probes both the top-layer O atoms (bright red in Fig. 1a) as well as the slightly lower O atoms located between the top zig-zag rows (dark red in Fig. 1a). In a LEIS spectrum taken after a single TPD experiment (as shown in Fig. 2a), the $^{16}O/^{18}O$ peak area ratio is ≈ 86/14. Because this signal comes from the top two O layers, this corresponds to ≈ 28 % of the top-layer O atoms being exchanged. The $^{18}O$ signal fraction decreased in subsequent scans due to the highly-focused ion beam causing sputter damage, thus the initial fraction of $^{18}$O is probably slightly higher. Consequently, the LEIS dataset is in quantitative agreement with the TPD data and we can safely conclude that ≈ 30 % of the top-layer lattice oxygen of the α-Fe$_2$O$_3$(1$\bar{1}$02) surface is replaced in a single adsorption/desorption cycle, and a whole top oxygen layer is replaced within minutes when exposed to ≈ 3×10$^{-8}$ mbar water vapor at 350 K.

**Experiments in liquid and at near-ambient conditions**

It is important to address the question whether a similar oxygen exchange process can happen when the sample is immersed in liquid water. To study this, we utilized the UHV-compatible liquid water dosing setup described in reference (34). Essentially, water vapor is condensed and frozen onto a liquid-nitrogen-cooled tip placed above a sample stage in a specially designed side chamber, which is separated from the main UHV chamber with a gate valve. Once an icicle forms, the chamber is evacuated by a cryosorption pump. Without breaking the UHV conditions, the sample is then introduced in the sample stage with the icicle still present (the vapor pressure of ice at cryogenic temperatures is very low). Then, the icicle is melted by heating the cooled tip, and the water drop falls onto the sample surface when it thaws. The chamber is subsequently re-evacuated, and the sample is transferred back to the main UHV chamber for analysis. This way, ultimate cleanliness of the liquid water exposure can be achieved (34-36).

In this set of experiments an $^{18}$O-labelled α-Fe$_2$O$_3$(1$\bar{1}$02) surface was prepared as a starting point, and the surface $^{18}$O/$^{16}$O ratio was determined after exposures to near-ambient-pressure H$_2$$^{16}$O vapor or liquid water. The $^{18}$O-labelling of the surface was achieved by annealing to 450 °C in a background pressure of 5×10$^{-7}$ mbar $^{18}$O$_2$, which consistently resulted in surface containing (61±2) % $^{18}$O, as measured by LEIS (Fig. 3a). In this experimental setup, the incidence direction of 1.225 keV He$^+$ beam is at an azimuthal angle of ≈ 105° from the direction of the surface rows, the polar angle is ≈ 45°. One would expect a much higher neutralization probability for the He$^+$ ions impinging the second layer oxygen (dark red in Fig. 1), because these pass very closely (≈ 1.0 – 1.6 Å) to the top-layer atoms. Indeed, this dataset shows a significantly higher Fe/O ratio compared to the previously shown experiments ((4.0±0.4) in this setup vs. (1.6±0.5) in the TPD chamber). Thus, the O signal is most likely dominated by the topmost O layer (bright red in Fig. 1).

Following the exposure to 6 mbar H$_2$$^{16}$O vapor and desorption of all adsorbed water molecules by heating to 450 K, the measured $^{18}$O fraction decreased by ≈ 1/3 to (40±3) % $^{18}$O; this result was identical in separate experiments with water vapor exposure time of 10 and 60 minutes, and also with seconds-long exposure to liquid H$_2$$^{16}$O (Fig. 3b). Longer exposure times to liquid water showed a somewhat higher exchange signal, but these experiments led to contamination by impurities washed from the sample mount (see Supplementary Figure 2). After these experiments, the XPS spectra show no significant change in the Fe 2$p$ region (Fig. 3d), but a small signal appears in the O 1$s$ and C 1$s$ regions (Fig. 3d,e). The whole spectrum is slightly shifted to higher binding energy, most likely due to adsorbate-induced band bending. In the C 1$s$ region, the small peak at ≈ 284 eV corresponds to "adventitious" carbon, while the ≈ 288 eV peak position is close to that observed for formate (HCOO$^-$) on similar oxide surfaces (35-37). As all the water should be desorbed after heating to 450 K, the ≈ 532 eV signal in the O 1$s$ region is assigned to the adsorbed carboxylic species. For quantifying the amount of carbonaceous contamination, the data are compared to reference spectra taken after dosing a saturation amount of formic acid on the surface at room temperature (gray dashed lines, possibly one full monolayer); in all ambient pressure experiments the carbonaceous signal was between 5 and 30 % of the saturation amount at room temperature. Importantly, the intensity of the carbonaceous signal does not correlate to the amount of exchanged oxygen observed in LEIS (see Supplementary Figure 2).



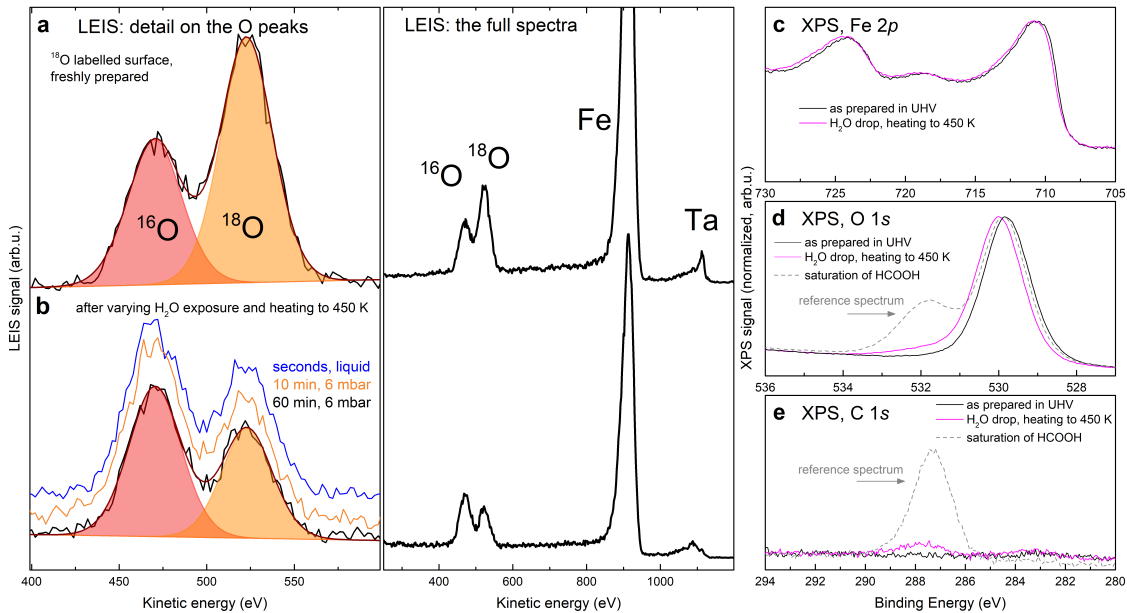

**Figure 3:** LEIS and XPS characterization of the α-Fe$_2$O$_3$(1$\bar{1}$02) surface before and after near-ambient pressure and liquid water exposures. (**a**) LEIS spectrum of the freshly prepared $^{18}$O-labelled surface. (**b**) LEIS spectra taken after prolonged exposure to 6 mbar H$_2$$^{16}$O and short liquid exposure show that ≈ 33 % of the probed O is exchanged. Spectra are offset for clarity. (**c-e**) XPS spectra (Mg Kα, 70° grazing emission) taken after surface preparation (black) and after 5 min. liquid water exposure (magenta). After the liquid water exposure a small C signal appears, related to carbonaceous contamination. From comparison to reference spectra (dashed gray) this carbonaceous signal amounts to 5 – 30 % of room temperature saturation of HCOO$^-$. Spectra shown in C,D are normalized to maximum, spectra shown in E are normalized to background. Source data are provided as a Source Data file.

These water-dosing experiments clearly demonstrate that a substantial amount of surface oxygen is exchanged already when the water exposure lasts only a few seconds. The experiments with varying exposure time between 10 and 60 minutes further show that the amount of exchanged oxygen stays constant, which provides a strong indication that no continuous exchange process takes place at these timescales. Assuming the top-layer sensitivity of LEIS in this geometry, the ≈ 1/3 decrease of the $^{18}$O signal agrees with the TPD and LEIS results acquired in the UHV-exposure experiments shown in Fig. 2. Thus, it is likely that the same mechanism takes place in both cases – after a submonolayer water dose in UHV and after a liquid water exposure.

**Density Functional Theory computations**

DFT computations help to understand the possible mechanisms of the oxygen exchange process (see Methods for details). Starting from the phase stable at room temperature (isolated HO-H$_2$O dimers, Fig. 1c), the computational results show that the first step is the desorption of the molecular H$_2$O. With the DFT functional employed, this H$_2$O desorption step costs 1.69 eV and results in the formation of isolated (HO-H) on the surface (red-marked panel in Fig. 4). Following this initial step, the remaining OH can either recombine with the neighboring H to form water and desorb at a cost of 1.38 eV (Direct desorption in Fig.4, black line), recombine to form water and diffuse along the (1$\bar{1}$0$\bar{1}$) direction (pathway A in Fig. 4, from the isolated OH-H phase to A6), or recombine to form water and diffuse along the $(11\bar{2}0)$ direction (pathway B in Fig. 4, from the isolated OH-H phase to B4). Both diffusion pathways are energetically cheaper than direct water desorption and eventually result in the recreation of another partially-dissociated HO-H$_2$O dimer in a nearby unit cell (panels A6 and B4). In pathway A, this is achieved by a rolling motion of the recombined H$_2$O along the zig-zag rows of the surface. Our computations reveal that the rate-limiting step of pathway A is the water molecule turning over, at which point it only binds to the surface via two hydrogen bonds (panel A1 in Fig. 4, barrier +0.80 eV). There are two such steps along the way to the HO-H$_2$O dimer recreation (A1 and A4), and both are slightly higher in energy compared to the rate-limiting step of pathway B (although the difference lies within the error of the computational setup). Pathway B involves the oxygen exchange with the lattice, and the rate-limiting step of this pathway is the formation of an HO-HOH-OH complex (panel B2, +0.73 eV) which contains an oxygen atom extracted from the surface. The relative stability of this complex stems from the presence of hydrogen and surface bonds stabilized by cooperativity effects. Such phenomena have been thoroughly studied on water/metal systems (38) and their validity has recently been demonstrated also on water/oxide surfaces (25, 39-42). Thus, the energy gain from cooperatively-strengthened bonds partially compensates the energetic cost of the lattice oxygen extraction. After the HO-HOH-OH complex is formed, the dissociated OH falls into the surface oxygen vacancy and the H$_2$O containing the extracted oxygen moves over to form the preferred HO-H$_2$O dimer (panels B3 and B4). Animations of the two diffusion pathways and relevant structural files are provided in the Supplementary Materials.

Our calculations at 0 K show that the diffusion is thermodynamically preferred over desorption by 0.65 eV, while the relative energetic difference between the considered diffusion pathways lie within the error of the computational setup: Pathway B is marginally preferred by 0.07 eV and exhibits the fastest diffusion rate (the onset of significant diffusion occurs at a temperature at least 25 K lower compared to pathway A, see Supplementary Note 5). Zero-point energy corrections and vibrational entropy effects on rates have been investigated, and do not affect the conclusions shown here (further details are provided in the Supplementary Note 4). From the HO-H$_2$O dimers re-formed by any of the considered pathways, the H$_2$O will eventually desorb and the remaining isolated (HO-H) will diffuse further. A plausible mechanism of oxygen exchange thus includes both diffusion pathways: Pathway B is available when there are neighboring (HO-H) in the $(11\bar{2}0)$ direction, whereas pathway A will take place in the absence of such an arrangement. Eventually, this leads to a significant fraction of the desorbing water containing an oxygen atom extracted from the α-Fe$_2$O$_3$ lattice.

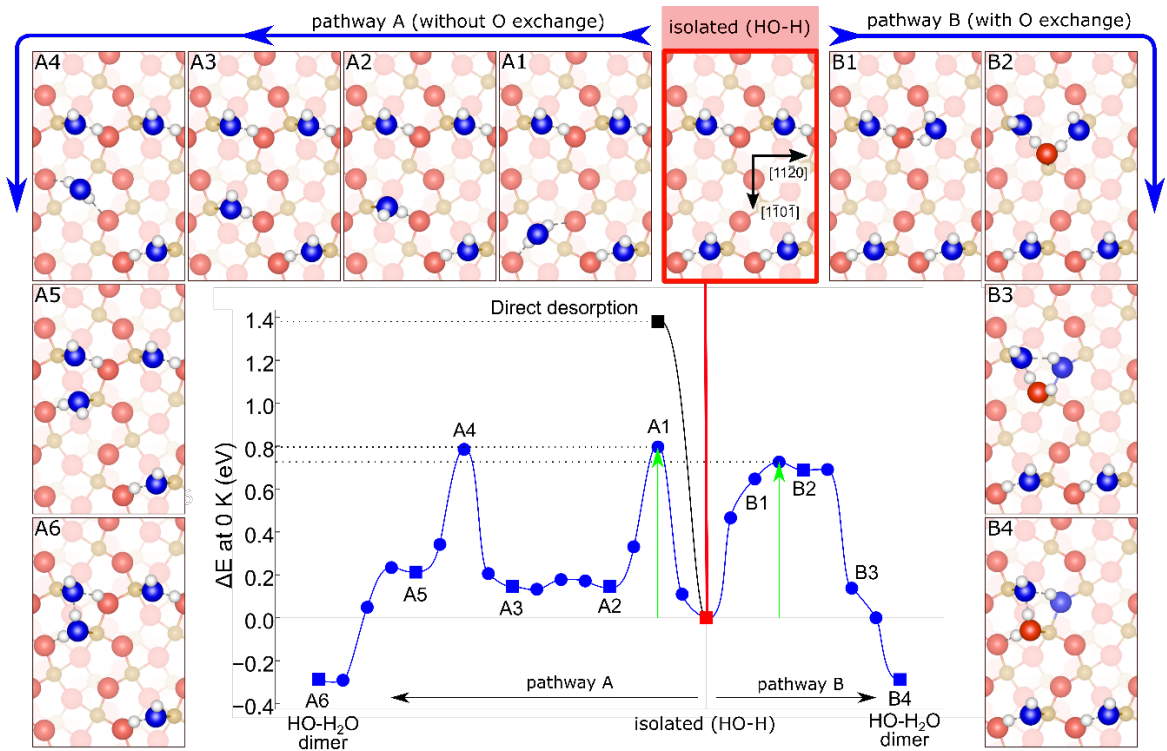

**Figure 4:** Minimum energy paths, obtained at a DFT level (0 K), for various $H_2O$ diffusion pathways on the α-$Fe_2O_3$($1\bar{1}02$) surface. Oxygen atoms originating from $H_2O$ are drawn in blue, oxygen atoms originating from the lattice are red, iron atoms are yellow. Pathway A features $H_2O$ diffusion along the zig-zag rows of the surface, pathway B features diffusion in the perpendicular direction. Both the diffusion mechanisms are close in energy and result in recreation of the HO-$H_2O$ water dimers; pathway B involves O exchange between the water and the lattice. Energetic cost of direct desorption is plotted for comparison (black). In the energy plot, squares correspond to the calculated configurations, connected by cl-NEB calculations shown as circles, with the highest point corresponding to the transition state. Rate-limiting steps are indicated by the green arrows. The transition state in path B is visually identical to B2 and therefore is not shown. Structure files and pathway animations are provided in the Supplementary Materials.

## Discussion

Our experimental observations clearly show that a large portion of the surface oxygen is exchanged when water molecules desorb from the α-$Fe_2O_3$($1\bar{1}02$) surface. This observation is highly unusual, as similar experiments performed on other well characterized metal oxide surfaces find that the oxygen exchange signal measured in similar conditions is typically very low and often restricted to point defects (19, 21, 43). Indeed, identical experiments on an apparently similar iron oxide surface, magnetite $Fe_3O_4$(001) (44), show the expected few percent exchange at > 500 K is linked to oxygen vacancies (43). The difference between these two systems is curious because both surfaces feature rows of 5-fold coordinated $Fe^{3+}$ cations alongside 3-fold coordinated $O^{2-}$, and water also adsorbs as partially dissociated HO-$H_2O$ dimers on $Fe_3O_4$(001) (39, 45). The key difference seems to be the particularly high adsorption energy for water on α-$Fe_2O_3$($1\bar{1}02$), which ensures that molecules remain on the surface at temperatures where extraction of lattice oxygen can occur. A second key ingredient clearly lies with the details of the surface structure, as our theoretical calculations find that cross-row diffusion through an OH-HOH-OH intermediate is the best explanation for our observations.

A particularly interesting aspect of our study is that the oxygen exchange occurs primarily during the diffusion of water at submonolayer coverages. The constant fraction of the exchanged signal after various near-ambient exposures further hints that no continuous exchange process is detectable at the experimental timescales (seconds to hours). This behavior is easily explained by a diffusion-exchange mechanism: Diffusion inherently requires unoccupied cation sites at the surface, and such a setting is naturally many orders of magnitude more likely to occur in vacuum than in ambient conditions. It is likely that a similar mechanism takes place in liquid on geological timescales, but this cannot be confirmed by a surface science experiment.

The OH-HOH-OH species that lie at the heart of our observations are stabilized by cooperativity effects between hydrogen and surface bonds (38, 39). The water molecules recombine or dissociate multiple times during the diffusion processes, which allows fulfilling the cooperativity rules in every reaction step. Partially-dissociated water agglomerates and networks are common on oxide surfaces (25, 39-42, 46, 47), but until now it could have been debated whether this has any real impact on surface reactivity. This example of facile oxygen exchange demonstrates that cooperativity is of paramount importance in surface chemistry processes involving water, which highlights the necessity to use adequate methodology to study solid-liquid interfaces. Crucial phenomena will be missed in classical molecular dynamics simulations using non-dissociable water molecules.

Atomic-scale identification of a rapid diffusion-exchange mechanism can be highly relevant for applications where control over atom exchange is desired, e.g. in catalysis or corrosion protection. Our work shows the oxygen extraction can be catalyzed by surface-bound water clusters, which allows rapid oxygen exchange even on surfaces where the thermodynamic cost of making an oxygen vacancy on a bare surface is high. Thus, identifying similar mechanisms on relevant surfaces and engineering systems exposing primarily exchange-resistant or exchange-allowing facets might be a viable approach, where a diffusion-exchange might be promoted by choosing conditions allowing incomplete occupancy of surface cation sites.

An interesting question is how relevant the rapid diffusion-exchange is for geochemistry and paleoclimatology, disciplines where the isotopic composition of water and minerals is of crucial importance (12, 48-50). Our study cannot imply much about slow processes happening in liquid at geological timescales, but it is relevant for conditions at which the geochemical samples are often prepared for analysis – i.e., when finely ground samples ($\approx$ 10 μm) are rinsed and dried at elevated temperatures in vacuum. Water vapor is normally the main component of residual gas; our study thus shows that one can expect the surface isotopic composition of a powder sample to be completely equilibrated with the surrounding vacuum within minutes. Of course, bulk diffusion would have to take place simultaneously to significantly affect the measured isotopic composition. Nevertheless, one needs to consider the ground powder already has a high surface-to-volume ratio, and room temperature bulk diffusion coefficients are often estimated by extrapolation from measurements at high temperatures, showing significant uncertainties (7, 51, 52). Thus, in our opinion it cannot be excluded that surface diffusion-exchange mechanisms might be contributing to systematic errors in mineral isotopic composition measurements.

In summary, we have elucidated the atomic-scale details of a surprisingly rapid oxygen exchange mechanism on a prototypical mineral surface. We have also explained why it is observed in vacuum studies but happens much slower (if at all) in liquid. Since no single parameter of the stoichiometric α-$Fe_2O_3$($1\bar{1}02$) surface is particularly unique, it seems reasonable to assume that many other oxide surfaces will permit similarly rapid oxygen exchange with water.

**Methods**

The experimental results were acquired in two independent UHV systems. The TPD and LEIS results following water vapor exposure in UHV were acquired in a chamber specifically designed for surface chemistry studies of oxide single crystals. This chamber has a base pressure of $8\times10^{-11}$ mbar and features liquid-He flow cryostat, a home-built calibrated molecular beam source, a HIDEN HAL 3F PIC quadrupole mass spectrometer, a focused ion gun with deflection unit (SPECS IQE 12/38), a hemispherical analyzer (SPECS Phoibos 150), a monochromated Al/Ag Kα X-Ray Source (SPECS XR50 M, FOCUS 500) and low energy electron diffraction optics (SPECS ErLEED). A natural α-$Fe_2O_3$($1\bar{1}02$) sample (SurfaceNet GmbH) is mounted on a Ta sample plate using Ta clips. A thin Au foil is put between the sample and the sample plate to ensure good thermal contact. The temperature is measured by a K-type thermocouple spot-welded on the sample plate; heating is provided by direct current. Precise gas dosing is done via an effusive molecular beam, which is formed by expansion of 0.53 mbar of high-purity $D_2O$ or $H_2^{18}O$ gas through two differentially-pumped stages. This results in a molecular beam with a known flux and a top-hat intensity profile, which under normal incidence results in a circular beam spot on the sample surface with a diameter of (3.32 ± 0.15) mm. Full details of this experimental system are provided in reference (53).

The near-ambient pressure and liquid exposure experiments were carried out in a custom-built compartment attached to a UHV chamber with a base pressure of $10^{-10}$ mbar. The liquid-dosing compartment is briefly described in the main text, full details are provided elsewhere (34, 35). The ambient pressure exposure experiments were carried out in the same setup without the use of the cold finger. For these experiments, the water reservoir was kept at 0 °C by cooling it with $LN_2$ to freeze some of the water and then allowing the bath temperature to equilibrate. This sets the vapor pressure to 6 mbar (54). The UHV chamber includes an ion source with deflection unit (SPECS IQE 12/38), a hemispherical analyzer (SPECS Phoibos 100), a Mg/Al Kα X-Ray source (VG XR3E2), a low energy electron diffraction setup (Omicron SPECTALEED) and a scanning tunneling microscope (Omicron STM-1).

The DFT calculations were carried out using the Vienna ab initio simulation package (55, 56), employing the projector augmented wave method (57, 58), with the plane-wave basis set cutoff energy set to 550 eV. Calculations are spin polarized and performed at the Γ-point for the 2×3 supercells used to study various diffusion mechanisms. Convergence is achieved when the electronic energy step of $10^{-6}$ eV is obtained, and forces acting on ions become smaller than 0.02 eV/Å. Diffusion activation energies are calculated using the nudged elastic band - climbing image (cl-NEB) method (59). The Perdew-Burke-Ernzerhof (PBE) (60) functional is used, with dispersion effects treated by Grimme's D2 method (61) for all calculations presented here. More advanced functionals have been tested, such as optB88-DF (62-64), but did not affect our conclusions regarding relative stabilities of configurations. Concerning NEB calculations, optB88-DF tends to get stuck in local minima and we have avoided its use. All results in this work are therefore at the D2 level and we only rely on a relative energy comparison.

An effective on-site Coulomb repulsion term $U_{eff}$ = 5, according to Dudarev et al. (65) was applied for the 3d electrons of the Fe atoms (similar results were obtained with a reduced $U_{eff}$ of 4 eV, as summarized in Supplementary Table 1). Symmetric slabs have been built, consisting of 4 $Fe_4O_6$ layers in thickness (256 atoms, 2×3 supercell) where only the 2 inner central O layers are kept fixed. The bottom surface is saturated with a full monolayer of water molecules and left untouched throughout the study. A void of 15 Å between consecutive slabs normal to the surface is added to avoid interactions. Zero point energies have been included using phonon density of states calculated with finite difference and the phonopy package (66). For the calculation of the diffusion rates, we have used several different computational methods including assuming the common prefactor value of $10^{13}$ s$^{-1}$, including or excluding VdW and ZPE corrections, using vibrational entropies instead of the assumed prefactor value, and correcting for the inconsistency between the modes of the initial and transition states. Details of these reaction rate calculations are provided in Supplementary Note 5. Further relevant information regarding the α-$Fe_2O_3$($1\bar{1}02$) surface can be found in our previous work (24, 25).

**Data availability:** All data needed to evaluate the conclusions in the paper are present in the paper and/or the Supplementary Information. Source data are provided as Source Data files.

**Acknowledgments:** GSP, ZJ and MM acknowledge funding from the Austrian Science Foundation (FWF) Start Prize Y847-N20. UD, FK and JB acknowledge the Austrian Science Fund FWF (Project 'Wittgenstein Prize, Z250-N27). GSP and MM acknowledge funding from the European Research Council (ERC) under the European Union's HORIZON2020 Research and Innovation program (ERC Grant Agreement No. [864628]). UD acknowledges support from the European Research Council, ERC-ADG 883395 WatFun. ZJ also acknowledges support from the TU Wien Doctoral Colleges TU-D. The computational results were achieved in part using the Vienna Scientific Cluster (VSC 3 and VSC 4)

**Author Contributions:** ZJ, FK and JB performed the experiments under the supervision of GSP, who conceptualized the research and acquired research funding for the project. MM performed the theoretical calculations under the supervision of CF. JB and JP designed some of the experimental equipment used. ZJ wrote the paper with substantial input/revision from MM, GSP, UD, MS, FK, CF.

**Competing Interest Statement:** Authors declare no competing interests.

**Figure Captions:**

**Figure 5:** Water adsorption on α-$Fe_2O_3$($1\bar{1}02$). (**a**) Top view on the bulk-truncated α-$Fe_2O_3$($1\bar{1}02$) surface. The red (larger) and yellow (smaller) balls correspond to oxygen and iron; the surface unit cell contains two O and two Fe atoms. A full analysis of this surface termination is provided in reference (24). In this work, the azimuthal angle of 0° is defined as the ($1\bar{1}0\bar{1}$) direction. (**b,c**) Two phases of water adsorbed on this surface, both of which consist of partially-dissociated water dimers (25). The bright blue and dark blue balls correspond to oxygen in $H_2O$ and OH, respectively. (**d**) $D_2O$ TPD (temperature ramp 1 K/s) shows two main desorption peaks (β, γ) before the multilayer forms (α). The saturation coverages of the β, γ peaks correspond to models shown in B and C, respectively.

**Figure 6:** Observation of the oxygen exchange between $H_2^{18}O$ and α-$Fe_2^{16}O_3$ by TPD and LEIS. (**a**) TPD spectrum (1 K/s) after a dose of ≈ 1.9 $H_2^{18}O$/u.c. at 120 K. The β peak is observed mostly in the $m/e$ = 20 signal, while the γ peak desorbs in two channels of similar intensity, indicating substantial oxygen exchange with the α-$Fe_2^{16}O_3$ substrate. (**b**) Oxygen exchange observed during 2.7×10$^{-8}$ mbar $H_2^{18}O$ exposure at 350 K. Upon opening of the molecular beam shutter, roughly 1/3 of the signal is observed in the $m/e$ = 18 channel. After ca. 360 s, almost all the signal is contained in the $m/e$ = 20 channel due to the saturation of the surface with $^{18}O$. The overall water signal remains constant. (**c**) LEIS characterization (1 keV He$^+$, scattering angle 90°) of the α-$Fe_2^{16}O_3$ as prepared (top), after the $H_2^{18}O$ exposure as shown in B (middle) and after a single $H_2^{18}O$ TPD experiment (bottom). The latter two spectra show substantial enrichment of the α-$Fe_2^{16}O_3$ surface with $^{18}O$. Source data are provided as a Source Data file.

**Figure 7:** LEIS and XPS characterization of the α-$Fe_2O_3$(1$\bar{1}$02) surface before and after near-ambient pressure and liquid water exposures. (**a**) LEIS spectrum of the freshly prepared $^{18}$O-labelled surface. (**b**) LEIS spectra taken after prolonged exposure to 6 mbar $H_2^{16}O$ and short liquid exposure show that ≈ 33 % of the probed O is exchanged. Spectra are offset for clarity. (**c-e**) XPS spectra (Mg Kα, 70° grazing emission) taken after surface preparation (black) and after 5 min. liquid water exposure (magenta). After the liquid water exposure a small C signal appears, related to carbonaceous contamination. From comparison to reference spectra (dashed gray) this carbonaceous signal amounts to 5 – 30 % of room temperature saturation of $HCOO^-$. Spectra shown in C,D are normalized to maximum, spectra shown in E are normalized to background. Source data are provided as a Source Data file.

**Figure 8:** Minimum energy paths, obtained at a DFT level (0 K), for various $H_2O$ diffusion pathways on the α-$Fe_2O_3$(1$\bar{1}$02) surface. Oxygen atoms originating from $H_2O$ are drawn in blue, oxygen atoms originating from the lattice are red, iron atoms are yellow. Pathway A features $H_2O$ diffusion along the zig-zag rows of the surface, pathway B features diffusion in the perpendicular direction. Both the diffusion mechanisms are close in energy and result in recreation of the HO-$H_2O$ water dimers; pathway B involves O exchange between the water and the lattice. Energetic cost of direct desorption is plotted for comparison (black). In the energy plot, squares correspond to the calculated configurations, connected by cl-NEB calculations shown as circles, with the highest point corresponding to the transition state. Rate-limiting steps are indicated by the green arrows. The transition state in path B is visually identical to B2 and therefore is not shown. Structure files and pathway animations are provided in the Supplementary Materials.

# Supplementary Information

## Rapid oxygen exchange between hematite and water vapor

**Authors**

Zdenek Jakub[1]†, Matthias Meier[1,2], Florian Kraushofer[1], Jan Balajka[1], Jiri Pavelec[1], Michael Schmid[1], Cesare Franchini[2,3], Ulrike Diebold[1], Gareth S. Parkinson[1*]

**Affiliations**

[1]Insitute of Applied Physics, TU Wien, Vienna, Austria

[2]University of Vienna, Faculty of Physics and Center for Computational Materials Science, Vienna, Austria

[3] Alma Mater Studiorum - Università di Bologna, Bologna, Italy

†current affiliation: Central European Institute of Technology (CEITEC), Brno University of Technology, Czech Republic

*correspondence to: parkinson@iap.tuwien.ac.at

## Supplementary Note 1

### Corrections of the *m/e* = 18 signal in the TPD experiments

The TPD data shown in Fig. 2A,B of the main text are corrected for water ($H_2^{16}O$) adsorption from the background, the cracking pattern of the $H_2^{18}O$ and the $H_2^{16}O$ impurity in the $H_2^{18}O$ beam. The correction of the *m/e* = 18 signal is calculated by the following formula:

$$I_{18,\text{corrected}} = I_{18,\text{raw}} - f_{\text{beam+crack}} * I_{20} - I_{18,\text{blank}},$$

where $I_{18,\text{raw}}$ is the raw data of the TPD scan (after subtraction of a constant offset due to higher *m/e* = 18 background), $f_{\text{beam+crack}}$ is the correction factor for the $H_2^{16}O$ signal in the $H_2^{18}O$ beam, $I_{20}$ is the *m/e* = 20 signal acquired simultaneously to $I_{18,\text{raw}}$ and $I_{18,\text{blank}}$ is the *m/e* = 18 signal measured in a blank experiment with no $H_2^{18}O$ dose (also with subtracted constant offset).

The plot of $I_{18,\text{raw}}$, $I_{20}$ and the smoothed $I_{18,\text{blank}}$ signal is shown in Supplementary Figure 1A. The blank experiment was conducted in the same arrangement with the same sample treatment and preparation, just without the actual $H_2^{18}O$ dose. The nonzero *m/e* = 18 signal in a blank experiment is due to the background adsorption of $H_2^{16}O$ on the sample and the sample mount. The background is increased (in the $10^{-10}$ mbar range) prior to the TPD experiment due to the previous annealing of the sample in $^{16}O_2$ background, which is known to displace $H_2O$ from the chamber walls.

The beam correction factor, $f_{beam+crack}$, accounts for the $m/e = 18$ signal coming from the $H_2^{18}O$ beam. It was determined from an experiment in which the $H_2^{18}O$ beam was directed into the chamber with the sample moved away, but the QMS ionizer moved near the beam trajectory. From this data, shown in Supplementary Figure 1B, it can be determined how much of the $m/e = 18$ signal is due to the cracking of $H_2^{18}O$ and $H_2^{16}O$ impurity in the beam. The $f_{beam+crack}$ was calculated as the ratio of the ($m/e = 18$) and ($m/e = 20$) signals averaged over the time when the beam shutter was opened (between 50 - 120 s in Fig. S1B). Depending on the constant background subtraction method, the value of $f_{beam+crack}$ lies between 0.11 (when signal minimum is taken as the background) and 0.05 (when the signal average before the beam shutter opening, 0 - 40 s, is taken as the background). For the data correction shown in the main text, a mean value of $f_{beam+crack} = 0.08$ was used.

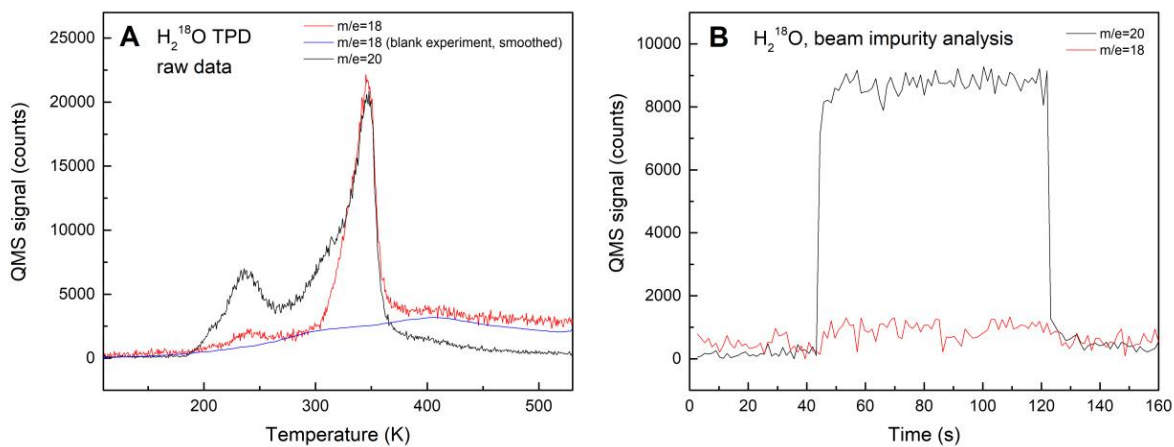

**Supplementary Figure 1: Corrections of the $m/e = 18$ signal.** A) Raw $H_2^{18}O$ TPD data showing simultaneously measured $m/e = 20$ and $m/e = 18$ signals. The blue curve shows an $m/e = 18$ signal measured in a blank experiment. In the processed data, this signal is subtracted from the $m/e = 18$ signal. B) Opening the $H_2^{18}O$ beam shutter into the mass spectrometer without the sample in sight shows the ratios of $m/e = 20$ and $m/e = 18$ signal. This is then used to correct the TPD signal for the cracking pattern and beam impurity. Source data are provided as a Source Data file.

# Supplementary Note 2

## Summary of all near-ambient-pressure and liquid-water experiments

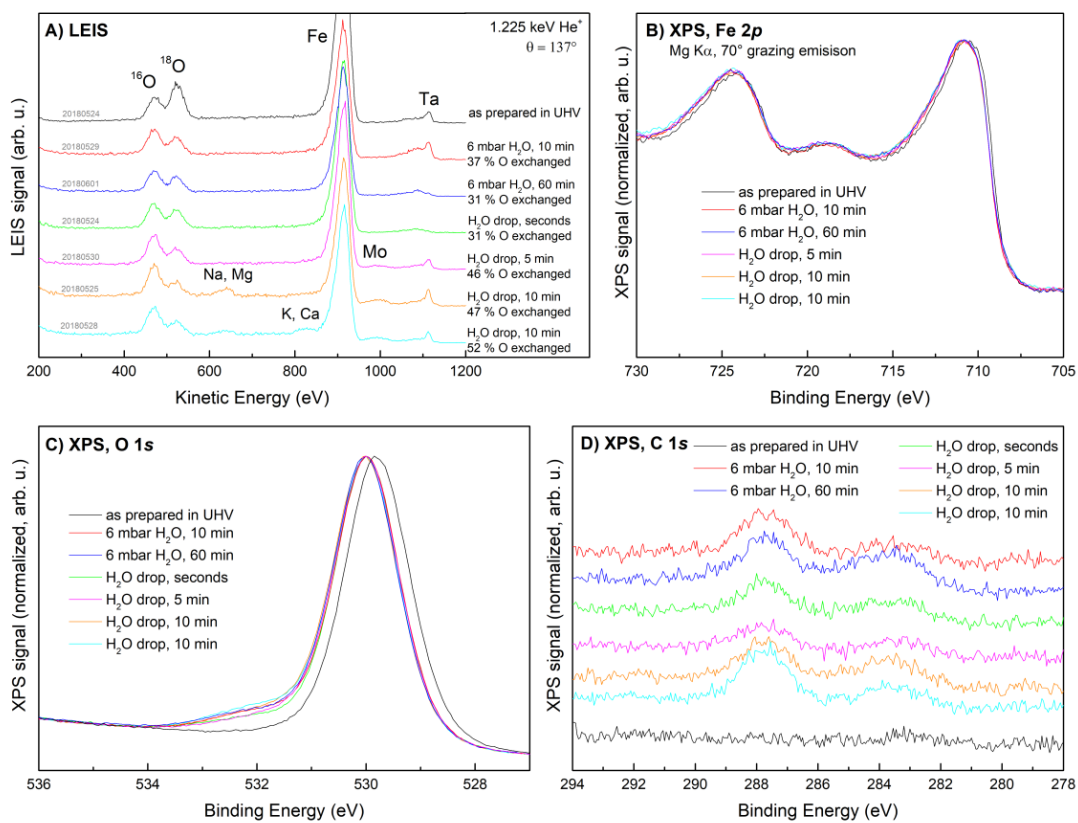

**Supplementary Figure 2: LEIS and XPS characterization of all the near-ambient pressure and liquid-water exposure experiments carried out within this project.** The LEIS spectra after prolonged liquid exposure (5 and 10 min) show higher fraction of $^{16}O$, but also show additional peaks whose position is consistent with Na, Mg, K, Ca and Mo. XPS spectra of all these experiments show small carbonaceous signal and an O 1s shoulder attributed to carboxylic species. Importantly, the amount of carbonaceous contamination signal observed in the C 1s region does not correlate to the amount of exchanged O observed in LEIS. Source data are provided as a Source Data file.

# Supplementary Note 3

### Animations of the considered diffusion pathways

Supplementary Movies 1 and 2 show visualizations of the two considered surface diffusion pathways of water molecules on α-$Fe_2O_3$(1$\bar{1}$02), labelled pathway A and pathway B in the main text. The color code is kept identical to the main text figure. For higher clarity, the surface atoms positions are kept constant. Adsorbates positions are according to the calculations, passing through NEB images, connected via a linear interpolation.

### Models of structures along pathways A and B

All the structures shown in Figure 4 in the main text are included as .cif files in the Supplementary Data Files.

## Supplementary Note 4

### The effect of $U_{eff}$ value and ZPE corrections on the computational results

The energies of transition states of pathways A and B computed using different values of $U_{eff}$ and considering different corrections are listed in Supplementary Table 1. The main conclusions are not affected by the varying methodology.

**Supplementary Table 1:** Transition state energies (eV) calculated with different $U_{eff}$ values and different corrections (vdW, ZPE).

|  | $U_{eff}=4$ D2-VdW | $U_{eff} = 5$ D2-VdW | $U_{eff} = 5$ no-VdW | $U_{eff} = 5$ no-VdW with ZPE |
|---|---|---|---|---|
| TS path A | 0.780 | 0.796 | 0.880 | 0.891 |
| TS path B | 0.707 | 0.726 | 0.688 | 0.671 |
| $TS^A$-$TS^B$ | 0.072 | 0.070 | 0.192 | 0.221 |
| 0 → B4/A6 | -0.267 | -0.288 | -0.228 | -0.168 |

## Supplementary Note 5

### Reaction rates of pathways A and B

Supplementary Figure 3 shows plots of reaction rates of pathways A and B using various computational methods. Comparison of panels A and B reveals that the exact choice of functional and treating of van der Waals interactions has a significant impact on the calculated rates, but it does not change the main conclusion that pathway B is preferred. Panel C shows rates calculated taking into account ZPE corrections of the relevant surface and adsorbate atoms involved in the diffusion process. By comparison to panel B, the ZPE corrections have surprisingly little effect on the calculated rates. The exchanged $O_{surf}$, lifted out of the surface plane shows similar changes in ZPE compared to $O_{mol}$ atoms of molecules creating additional bonds when forming dimers. Panel D shows rates calculated using the vibrational entropies instead of the commonly assumed prefactor value of $10^{13}$ s$^{-1}$. Panel E is included for completeness, as due to transition states having only 3N−1 modes, calculating ΔS entities leads to a small error. To take this into account we added a 1D translational mode to the transition state. The characteristic length of choice was 5 Å and leads to a shift to higher temperature. The quality of such an approximation can be discussed, but doesn't affect our conclusions, as both mechanisms are of the same nature.

Overall, all the tested methods lead to the same conclusion that pathway B is preferred over pathway A.

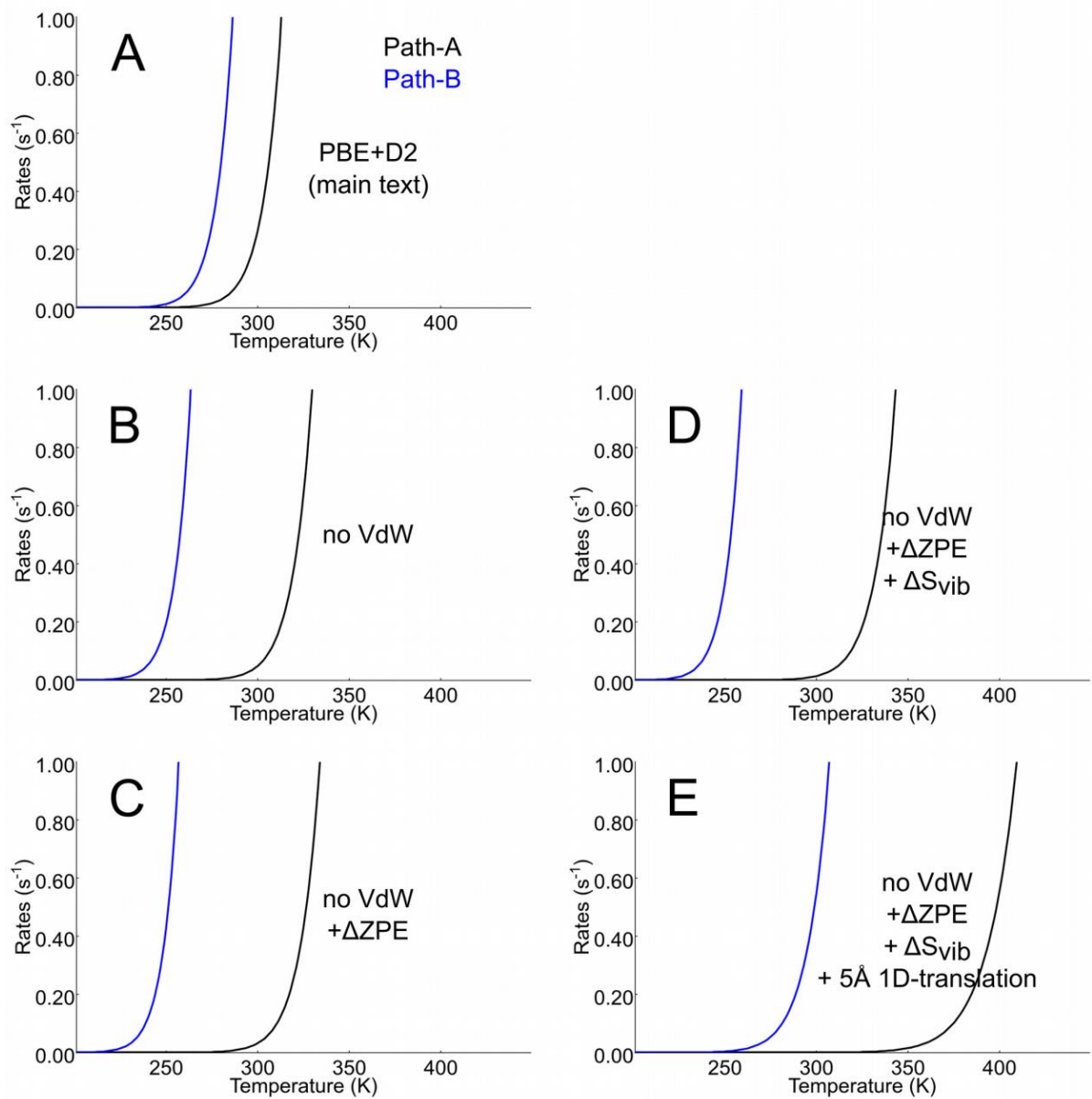

**Supplementary Figure 3: Comparison of reaction rates for pathway A (black) and B (blue) using different assumptions/approximations.** The rates are calculated: (A) assuming a prefactor of $10^{13}\,s^{-1}$ ($\Delta S = 0$) and using the PBE+D2 energies from the main text, (B) removing the VdW corrections, (C) adding ZPE corrections. (D) The respective vibrational entropies are estimated and used instead of the previous assumptions of $\Delta S=0$, in addition to the ZPE corrections. (E) Due to the inconsistency regarding total modes between initial and transition state (3N vs 3N−1 modes), we added a 1D translational contribution to the transition state (entropy and enthalpy components), with a characteristic length of 5 Å. In all the tested cases, pathway B is preferred over pathway A.